\documentstyle[eqsecnum,preprint,aps]{revtex}
\tightenlines

\def\pacscopy#1#2{\par %
\bgroup
\hsize\columnwidth \parindent0pt
\if@twocolumn\else\leftskip=0.10753\textwidth \rightskip\leftskip\fi
\ifdim\prevdepth=-1000pt \prevdepth0pt\fi
\dimen0=-\prevdepth \advance\dimen0 by20pt\nointerlineskip
\vbox to28pt{\small\vrule height\dimen0 width0pt\relax\ifdraft#1\fi\vfill}%
\egroup
\if@twocolumn\vskip1pc\fi
\ifpreprintsty
\penalty10000\vfill
\hbox to\columnwidth{\copyright#2\hfil Typeset using {REV\TeX}}\newpage
\fi
}

\def\beq{\begin{equation}}
\def\eeq{\end{equation}}
\def\bea{\begin{eqnarray}}
\def\eea{\end{eqnarray}}
\def\nnu{\nonumber}

\def\tta{\theta}

\def\om{\omega}
\def\lam{\lambda}

\def\Gam{\Gamma}
\def\Dta{\Delta}

\def\sech{{\rm sech\,}}
\def\ptl{\partial}

\def\zhat{\bf{\hat z}}

\def\lp{\left(}
\def\rp{\right)}

\def\ham{{\cal H}}
\def\ket#1{|#1\rangle}

\def\mel#1#2#3{\langle#1|#2|#3\rangle}

\def\bJ{{\bf J}}
\def\bH{{\bf H}}
\def\faca{(1-u_0^2)}
\def\facb{(1-u_0^2 - \lam)}
\def\hafa{\faca^{1/2}}
\def\hafb{\facb^{1/2}}
\def\hafc{(u_0^2 + \lam -1)^{1/2}}
\def\rta{\sqrt{1-u_0^2}}

\def\rtl{\sqrt\lam}
\def\rtlb{\sqrt{1 - \lam}}
\def\rtd{\sqrt{(1-u_0^2)(1-\lam)}}

\def\D0{\Delta_0}

\def\e0k{E_0^{(k)}}

\begin{document}
\draft

\title{Large transverse field tunnel splittings in the Fe$_8$ spin Hamiltonian}

\author{Anupam Garg}
\address{Department of Physics and Astronomy, Northwestern University,
Evanston, Illinois 60208}

\date{\today}

\maketitle

\begin{abstract}
The spin Hamiltonian that describes the molecular magnet Fe$_8$ has
biaxial symmetry with mutually perpendicular easy, medium, and hard
magnetic axes. Previous calculations of the ground state tunnel splittings
in the presence of a magnetic field along the hard axis are extended,
and the meaning of the previously discovered oscillation of this
splitting is further clarified.\hfill
\end{abstract}

\pacscopy{75.50.Xx, 75.45.+j, 03.65.Sq, 75.10.Dg}{1999, The American Physical Society}

\widetext

\section{Introduction}
\label{Intro}

The molecular magnetic cluster [(tacn)$_6$Fe$_8$O$_2$(OH)$_{12}$]$^{8+}$
(henceforth abbreviated to Fe$_8$) is approximately described
by a spin Hamiltonian of the type\cite{alb,sop,cam}
\beq
\ham  =  k_1 J_z^2 + k_2 J_y^2 - g\mu_B \bJ\cdot\bH, \label{ham} 
\eeq
where $\bJ$ is a dimensionless spin operator, $k_1 > k_2 > 0$, and $\bH$ is an
externally applied magnetic field. For Fe$_8$, $J=10$,
$k_1 \approx 0.33$ K, and $k_2 \approx 0.22$ K. The zero-field Hamiltonian
has biaxial symmetry with easy, medium, and hard axes along $x$, $y$, and
$z$ respectively \cite{fn1}.

The magnetization dynamics of crystals of Fe$_8$ and of a similar
compound, Mn$_{12}$-acetate, show some remarkable
phenomena\cite{sop,fs,tlbd},
which are not fully understood, and in this author's
view require a consideration of the spin-phonon interaction in addition
to the pure spin Hamiltonian (\ref{ham})~\cite{vhsr,agMn}. It is not
the purpose of this paper to discuss these issues. It seems that at least
some aspects of experimental behaviour~\cite{ws}
can be understood on the basis of
the spin Hamiltonian alone, and it is therefore worthwhile to address
the simpler one-particle-quantum-mechanics problem of understanding the
eigenvalue spectrum of Eq.~(\ref{ham}). Several authors have taken this
position, and that is our motivation in this paper too.

In particular, we wish to focus on the behaviour of the two lowest energy
levels when the external magnetic field $\bH$ is along the hard axis, $z$.
It was predicted in Ref.~\cite{agepl} that the tunnel splitting between
these two levels would oscillate as a function of the field strength, and
these oscillations appear to have been seen by Wernsdorfer and
Sessoli~\cite{ws,fn1}. Very briefly, for
$H < H_c = 2k_1 J/g\mu_B$, the spin system has two classical ground states,
in which the spin is canted
away from the $\pm x$ directions toward $z$. The classical degeneracy of
these states is lifted by quantum tunneling. The tunnel splitting, $\Dta$,
oscillates as a function of $H$, vanishing at $2J$ field
values lying in the interval $(-H^*,H^*)$, where
\beq
H^* = (1-\lam)^{1/2} H_c, \label{Hst}
\eeq
and $\lam = k_2/k_1$. (This quantity, along with other important parameter
combinations, is tabulated in Table I.) The vanishing fields are located
symmetrically around $H=0$; thus, in the case of half-integral $J$, 
$H=0$ is a vanishing point, in accord with Kramers' theorem.

The above calculation was done using instanton methods, wherein the
oscillation in $\Dta$ arises from the presence of two instantons, both of
which have a complex Euclidean action. The imaginary parts of this action are
equal and opposite for the two instantons (in a suitable gauge), and the
vanishing of $\Dta$ can be understood as due to superposition or
destructive interference between the corresponding amplitudes~\cite{ldg,vdh}.

Recently, Chudnovsky and Hidalgo~\cite{ch} have examined this problem for
$H > H^*$, and noted that in this field range the splitting grows
monotonically instead of oscillating. This is said by them to be a new
topological effect or quantum interference phenomenon, arising from a
new type of instanton. Indeed, their calculation of the splitting is quite
intricate, and the imaginary part of the action vanishes due to a
cancellation of two seemingly unrelated integrals, which they justifiably
refer to as remarkable.

In fact, the monotonicity of $\Dta$ for $H > H^*$ is already known
and, we shall argue, entirely expected on physical grounds. This is done in
Sec.~II. We shall also
present another way of calculating $\Dta$, which is a simple
extension of that in Ref.~\cite{agepl}, requiring nothing more than careful
analytic continuation and contour integration. The imaginary part of
the action vanishes in a simple and unremarkable way in this calculation.
This is done in Sec.~III. 

\section{Physical reason for oscillation of tunnel splitting}
\label{reason}

As noted in Ref.~\cite{ag3}, there is another way to understand the
vanishing of the tunnel splitting. For
$\bH\parallel \zhat$, the Hamiltonian~(\ref{ham}) becomes
\beq
\ham  =  k_1 J_z^2 + k_2 J_y^2 - g\mu_B HJ_z, \label{hamz} 
\eeq
which is invariant under a 180$^\circ$ rotation about $\zhat$.
This symmetry is reflected in the selection rule 
$\mel{m}{\ham}{m\pm1} = 0$, where $\ket m$ is a 
$J_z$ eigenstate. The Hamiltonian divides into two disjoint subspaces,
$V_+$, spanned by $m  =  J$, $J-2$, $\ldots$,
and $V_-$, spanned by $m=J-1$, $J-3$, $\ldots$.
The lowest eigenvalues (among others) from different spaces can cross
as $H$ is varied. The vanishing of the tunnel splitting is no more
than such a level crossing. It was shown in Ref.~\cite{ag3} (see also Fig. 1
there) that there are exactly $2J$ level crossings as $H$ varies from
$-H_c$ to $H_c$. Since our previous instanton calculation~\cite{agepl}
finds just as many crossings in the smaller field range $(-H^*,H^*)$, it
follows that there cannot be any more crossings or oscillations in $\Dta$ for
$H > H^*$ (or $H < -H^*$). Chudnovsky and Hidalgo's findings merely reflect
this already known fact.

That $\Dta$ grows monotonically for large $H$ is also to be expected
on physical grounds. As the field grows and approaches $H_c$, the classical
ground state orientations approach closer to $\zhat$. The angle between
them decreases, as does the energy barrier bewteen them. The default
behavior in such a case is for the splitting to grow. It is the oscillation
in $\Dta$ which is surprising, not the monotonic growth.

The above argument shows that the vanishing of $\Dta$ is really not a
topological effect in that it is not robust against perturbations
such as a small
misalignment of the magnetic field. Nevertheless, the instanton method
provides an effective way of approximately calculating the
splitting and the crossing points. 

\section{Instanton calculation of splitting for $H>H^*$}
\label{calcn}

We now show how to adapt the calculation of the tunnel splitting $\Dta$
in Ref.~\cite{agepl}, where we only gave explicit results for $|H| < H^*$,
to the case $H > H^*$. We refer readers to our previous paper for
the calculational set-up. We also follow our previous notation, and
summarize the important parameters in Table I.

In the instanton method, the splitting is given by
\beq
\Dta =  \Bigl|\sum_i \om_i e^{-S_i} \Bigr|, \label{Dtagen}
\eeq
where $\om_i$ is a quantity with dimensions of frequency or energy, 
and $S_i$ is the Euclidean action along a semiclassical path, or
instanton,
connecting the two states between which one is tunneling. The sum runs
over all possible instantons. In simple problems there is only one instanton,
but the converse is possible, especially in cases with high symmetry. This
is the case for our problem, where there are two instantons.

The dominant behavior of $\Dta$ is controlled by the action $S$, so
we will focus on finding only this quantity.
For our problem, symmetry guarantees that the prefactors $\om_i$
are identical for the two instantons involved, and the vanishing or
nonvanishing of $\Dta$ hinges entirely on whether or not the actions
$S_i$ are real or complex, and if the latter, what the relative
phase of the corresponding amplitudes is. Thus there is no need to
find the prefactors explicitly.

The spin paths in question can be given in terms of spherical polar
coordinates $\left(\tta(\tau),\phi(\tau)\right)$ where $\tau$ is an
imaginary time. For a given path, the action is given by
\beq
S = \int \lp iJ(1 - \cos\tta) \dot\phi + E(\tta,\phi) \rp d\tau,
   \label{act1}
\eeq
where the dot denotes a $\tau$-derivative, and $E(\tta,\phi)$ is
the classical energy or expectation value of $\ham$ in the spin
coherent state that is maximally aligned along the direction
$(\tta,\phi)$. For Eq.~(\ref{hamz}), we get, up to an additive constant,
\beq
E(\tta,\phi) = k_1J^2 \left[
                (\cos\tta - \cos\tta_0)^2 +
                        \lam\sin^2\tta \sin^2\phi \right],
  \label{etp}
\eeq
where $\cos\tta_0 \equiv u_0 = H/H_c$, and $\lam = k_2/k_1$. Note that
$u_0$ and $\lam$ are both less than unity. This energy has minima at
$(\tta,\phi) = (\tta_0,0)$ and $(\tta_0,\pi)$, and the additive constant
has been adjusted to make $E=0$ at these minima.

The instanton paths are those for which the action is stationary, and
which thus obey the Euler-Lagrange equations,
\bea
iJ\dot\tta \sin\tta &=& {\ptl E \over \ptl\phi}
                      = 2\lam k_1 J^2 \sin^2\tta \sin\phi \cos\phi,
         \label{tdot} \\
iJ\dot\phi \sin\tta &=& -{\ptl E \over \ptl\tta} 
                      = 2k_1J^2 \sin\tta \left[
                          (\cos\tta - \cos\tta_0) -
                        \lam\cos\tta \sin^2\phi \right].
         \label{pdot}
\eea 
The boundary conditions are that the path approach the two minima
as $\tau\to \pm\infty$.

Along an instanton, $E$ is conserved, so that the orbit
without regard to its time dependence can be found purely by
using energy conservation. This point is of great utility in
calculating the instanton action, for if we arrange for $E$ to equal zero,
the action is given entirely by the first term in Eq.~(\ref{act1}),
which can be written as an integral over $\phi$:
\beq
S_{\rm instanton} = iJ \int \lp 1 - \cos\tta(\phi) \rp d\phi.
           \label{Sint}
\eeq
In this integral, it is no longer necessary to regard $\phi$ as running
over the original instanton contour given by $\phi(\tau)$. Any deformation
of this contour is allowed as long as it does not encounter any
singularities.
This was the procedure followed in Ref.~\cite{agepl}. Setting $E=0$ in
Eq.~(\ref{etp}) we obtain
\beq
\cos\tta = {u_0 + i\lam^{1/2} \sin\phi F(\phi) \over
                 1- \lam\sin^2\phi}, \label{orbit}
\eeq
where
\beq
F(\phi) = (1 - u_0^2 - \lam\sin^2\phi)^{1/2}.  \label{defF}
\eeq
It is important to note that which branch of the square root is taken
depends on whether the instanton starts (at $\tau = -\infty$) at
$\phi= 0$ or $\phi = \pi$. Let us consider the former case. 
It is obvious from symmetry that there are two instanton paths, 
$(\tta_\pm(\tau),\phi_\pm(\tau))$, which wind around $\zhat$ in
opposite senses. Examination of Eqs.~(\ref{tdot}) and (\ref{pdot}) shows
that we must take $F(\phi) > 0$ for {\it both} $+$ and $-$ instantons.
The calculation is direct in the case
$u_0 < (1 - \lam)^{1/2}$, which is equivalent to $H < H^*$, for
then, the integration contour for $\phi$ can be taken to run from
$0$ to $\pm \pi$ while
keeping $F(\phi)$ real. The integration in Eq.~(\ref{Sint}) yields
\beq
S_\pm = S_r \pm i S_i,
\eeq
where
\bea
S_r &=& J \left[
           \ln \lp {\rta + \rtl \over \rta - \rtl} \rp
           - {u_0 \over \rtlb}
              \ln \lp {\rtd + u_0 \rtl \over \rtd - u_0 \rtl} \rp
            \right],  \label{Sr} \\ 
S_i &=& J \pi ( 1 - u_0/\rtlb). \label{Sim}
\eea
The splitting is thus proportional to $\exp(-S_r) \cos(S_i)$, which
oscillates as $H$ is increased as discussed above. Note that
$S_i = 0$ when $u_0 = (1-\lam)^{1/2}$.

In the case when $u_0 > (1-\lam)^{1/2}$,
i.e., when $H > H^*$, one may guess the answer by reasoning that
it must be analytic in the magnetic field, and that there should be
no imaginary part to the action, since $\Dta$ grows with $H$ as
argued in Sec.~II. And in fact, we notice that $S_i$ is exactly what
one would obtain if the denominators of the arguments
of the logarithms in $S_r$ were multiplied by $-1$, and the branches of
the logarithm interpreted suitably. This leads us to write
\beq
S_\pm = J \left[
           \ln \lp {\rtl + \rta \over \rtl - \rta} \rp
           - {u_0 \over \rtlb}
              \ln \lp {u_0 \rtl + \rtd \over u_0 \rtl - \rtd} \rp
            \right].  \label{Sall}
\eeq
This expression is real when $H>H^*$. In the other case,
it can be made to agree with Eqs.~(\ref{Sr}) and (\ref{Sim}) if we
stipulate that for $x < 0$, $\ln x$ is to be taken as $\ln |x| \mp i\pi$
for $S_\pm$. It turns out that Eq.~(\ref{Sall}) {\it is} correct.

Some readers may find the above argument too glib and reasoned after the
fact. Is it possible to obtain it by honest calculation?
The immediate problem is that when $u_0 > (1-\lam)^{1/2}$,
$F(\phi)$ is imaginary for $\phi$ in the real interval
${\cal M} = (\phi_1, \pi -\phi_1)$, where $\sin^2\phi_1 = (1-u_0)^2/\lam$.
One can proceed just as in the low field case by using Eqs.~(\ref{Sint})
and (\ref{orbit}), keeping $\phi$ real. $\cos\tta$ is then purely real
in the interval $\cal M$, and the resulting imaginary contribution to the
action cancels that from outside $\cal M$. This cancellation has the same
unsatisfactory and accidental character as that in Chudnovsky and
Hidalgo's calculation \cite{ch}. Further, one cannot decide which branch of
$F(\phi)$ should be used in $\cal M$ solely from energy conservation.

Another procedure, which also entails simpler integrals, is to
find the $\tau$-dependence of the instanton, and evaluate $S$ as
an integral over $\tau$ via Eq.~(\ref{act1}). 
We record the relevant formulas for the case $H < H^*$ first. 
Only one instanton, say the
$+$ one, need be considered explicitly. Substituting Eq.~(\ref{orbit})
in Eq.~(\ref{pdot}) we obtain
\beq
\dot\phi = \om_0 \sin\phi F(\phi), \label{pdeq}
\eeq
where $\om_0 = 2k_1 J \lam^{1/2}$. This equation can be integrated easily.
Defining $z = \hafa\om_0\tau$, and
\beq
G(z) = (1 - u_0^2 - \lam\tanh^2 z)^{1/2}, \label{defG1}
\eeq
we obtain
\bea
\cos\phi &=&  - \hafb \tanh z / G(z), 
               \quad {\rm or}, \label{cosph}\\
\tanh z &=& -\hafa \cos\phi /F(\phi). \label{tanhz}
\eea
It is easily verified that $\phi\to 0,\ \pi$, as $\tau \to \pm\infty$.

We also obtain
\beq
{d \phi \over dz} = \hafa\hafb \sech z/G^2(z), \label{dpdz} 
\eeq
\beq
1 - \cos\tta = (1 - u_0)
                 {1 - u_0^2 - \lam - u_0\lam \sech^2z 
                             - i(1+u_0)\lam^{1/2}\hafb \sech z
                    \over
                  (1-u_0^2)(1-\lam) - u_0^2\lam \tanh^2 z}.
                                                    \label{ctta}
\eeq

When $u_0 > (1-\lam)^{1/2}$, it is clear that Eqs.~(\ref{pdeq}--\ref{ctta})
must continue to hold as general analytic relations since they arise from
integration of the equations of motion which are unchanged. The only
subtlety is in the assignment of branches to the various multivalued
quantities and functions involved. Properly speaking, the integral in
Eq.~(\ref{Sint}) should be taken over a $\phi$ contour which is suitably
indented around the branch points. It is clearly permissible to parametrize
this contour in terms of any other variable, and one choice which is
often convenient is to use the original time-like variable $\tau$, but
allow it to become complex. With this point in mind
we first make $F(\phi)$ analytic by using a two-sheeted $\phi$
plane with a branch cut joining the points $\phi$ and $\pi - \phi_1$.
We then integrate Eq.~(\ref{pdeq}) as
\beq
z = \int_{\Gam} {(1-u_0^2)^{1/2} \over \sin\phi F(\phi)} d\phi,
      \label{zint}
\eeq
where the contour $\Gam$ is shown in Fig.~1. We leave it to the reader to
verify that a consistent phase assignment can be made to $F(\phi)$ in
such a way that along this contour, we have
\beq
{\rm arg}\ F(\phi) = \cases{
   0 & $\phi < \phi_1$; $\phi > \pi - \phi_1$; \cr
   \pi/2 & $\phi_1 < \phi < \pi - \phi_1$. \cr}
\label{argF}
\eeq
The corresponding contour ${\cal C}$ traced out in the $z$ plane is
drawn in Fig.~2. The easiest way to see this is that for $\phi$ outside
the cut, the integrand in Eq.~(\ref{zint}) is real, so the corresponding
parts of ${\cal C}$ must be parallel to the real axis. Secondly, the
change in $z$ as $\phi$ runs across the cut is given by
\beq
\Dta z = e^{-i\pi/2} \int_{\phi_1}^{\pi - \phi_1}
             {\sin\phi_1 \over \sin\phi (\sin^2\phi - \sin^2\phi_1)^{1/2}}
              d\phi = -i\pi. \label{jump}
\eeq
Using the arbitrary constant of integration to adjust the overall vertical
position of ${\cal C}$, we arrive at Fig.~2. We do not bother
to ask how ${\cal C}$ is indented around the poles of $\tanh z$ at
$\pm i\pi/2$, because it turns out that the integrand for the action
integral is analytic at these points.

The problem is now reduced to integrating the product of the factors
$(1-\cos\tta)$ and $d\phi/dz$ over the contour ${\cal C}$. The factors 
are given by Eqs.~(\ref{dpdz}) and (\ref{ctta}), with the proviso
that
\beq
\hafb \to i \hafc. \label{hafbc}
\eeq
The correctness of this choice follows by noting that for $\phi = \pi/2$,
$\dot\phi = \om_0 \hafb$, which must be positive and imaginary given our
contour ${\cal C}$. Collecting together Eqs.~(\ref{hafbc}), (\ref{dpdz}),
(\ref{ctta}), and (\ref{act1}), we obtain
\bea
S_+ = &J& {\faca^{3/2}\hafc \over 1 + u_0} \nnu \\
      &\times& \int_{\cal C}
        {u_0^2 + \lam - 1 + u_0\lam\sech^2z 
                  - \lam^{1/2}\hafc (1+u_0) \sech z
           \over
          \lp(1-u_0^2)(1-\lam) - u_0^2\lam \tanh^2z \rp
          (1 - u_0^2 - \lam \tanh^2z)}
             \sech z\ dz.
\label{Spl2}
\eea
All square roots in this formula are regarded as positive.
We now introduce the definitions
\beq
\tanh^2z_1 = {(1-u_0^2)(1-\lam) \over u_0^2\lam}, \quad
\tanh^2z_2 = {1-u_0^2 \over \lam}, \label{z12}
\eeq
and the attendant results
\beq
\sech z_2 = u_0\sech z_1 = \lam^{-1/2} \hafc. \label{sech12}
\eeq
The result for $S_+$ simplifies greatly in terms of these quantities.
The apparent singularities at $z = \pm z_1$ and $z = \pm z_2$ are
cancelled by factors in the numerator, and we end up with
\beq
S_+ = J {\faca^{3/2}\hafc \over u_0 \lam (1+u_0)} \int_{\cal C}
        {\sech z \over (\sech z + \sech z_1)(\sech z + \sech z_2)} dz.
\eeq
The integrand now has no singularities in the strip
$ -\pi < {\rm Im}\ z < \pi$. The contour ${\cal C}$ may thus be
deformed into the real line, which proves that $S_+$ is purely real.
The actual integral itself is elementary, and may be done with the
help of Ref.~\cite{gr}. The final result is given by
Eq.~(\ref{Sall}). The same result is obtained for $S_-$.

\acknowledgments
I am indebted to Wolfgang Wernsdorfer for useful correspondence
and discussions.
This work is supported by the NSF via grant number DMR-9616749.

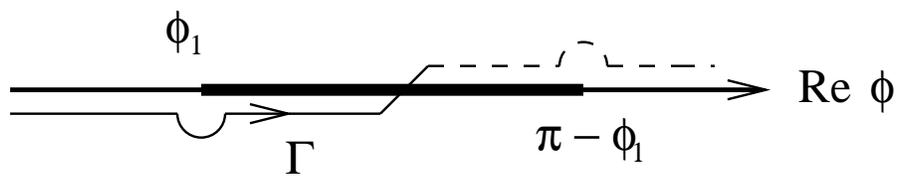
\begin{figure}
\caption{The integration contour $\Gam$ in the $\phi$ plane. The heavy
line denotes a branch cut, and the dashed
section of the contour lies on the second Riemann sheet.}
\end{figure}
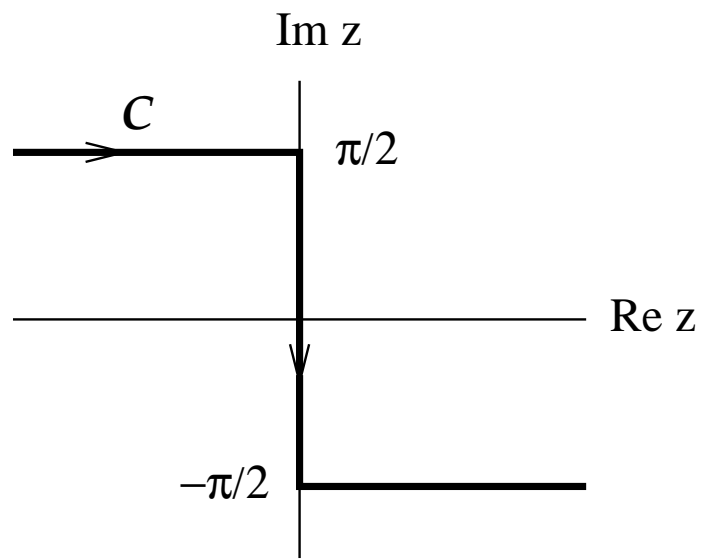
\begin{figure}
\caption{The contour ${\cal C}$ in the complex $z$ plane.}
\end{figure}

\begin{table}[t]
\caption{Summary of important parameter combinations}
\vspace{0.2cm}
\begin{center}
\begin{tabular}{c c}
Quantity & Formula \\
\hline
$H_c$                    &    $2k_1 J/g\mu_B$     \\
$u_0$, $\cos\tta_0$      &    $H/H_c$             \\
$\lam$                   &    $k_2/k_1$           \\
$H^*$                    &    $(1-\lam)^{1/2}H_c$ \\
$\om_0$                  &    $2J(k_1 k_2)^{1/2}$   \\
$\sin\phi_1$             &    $[(1-u_0^2)/\lam]^{1/2}$   \\
$z$                      &    $\hafa\om_0 \tau$
\end{tabular}
\end{center}
\end{table}
\end{document}